%% file: main.tex
\documentclass[a4paper,twoside]{article}

\usepackage{epsfig}
\usepackage{subcaption}
\usepackage{calc}
\usepackage{amsthm}
\usepackage{multicol}
\usepackage{pslatex}
\usepackage{apalike}
\usepackage[bottom]{footmisc}
\usepackage{amssymb}
\usepackage{amsthm}
\usepackage{amsmath}
\usepackage{stmaryrd}
\usepackage{listings}
\usepackage{multicol}
\usepackage{algorithm}
\usepackage{algpseudocode}
\usepackage{tikz}

\usetikzlibrary{matrix, positioning}
\usetikzlibrary{shapes,arrows}
\usepackage{graphicx} 
\usepackage{todonotes}
\usepackage{SCITEPRESS}     

\newtheorem{defin}{Definition}[section]

\newtheorem{lem}{Lemme}[section]

\newtheorem{rem}{Remark}[subsection]

\usepackage{enumitem}
\setitemize{noitemsep,topsep=0pt,parsep=0pt,partopsep=0pt}
\setlist[enumerate]{itemsep=0mm}

\algnewcommand{\IfThenElse}[3]{
  \State \algorithmicif\ #1\ \algorithmicthen\ #2\ \algorithmicelse\ #3}

\begin{document}

\title{Authentication Attacks on Projection-based  Cancelable Biometric Schemes}

\author{\authorname{DURBET Axel\sup{1}, GROLLEMUND Paul-Marie\sup{2}, LAFOURCADE Pascal\sup{1}, MIGDAL Denis\sup{1} and THIRY-ATIGHEHCHI\sup{1}}
    \affiliation{\sup{1}Université Clermont-Auvergne, CNRS, Mines de Saint-Étienne, LIMOS, France}
    \affiliation{\sup{2}Université Clermont-Auvergne, CNRS, LMBP, France}
}


\abstract{ Cancelable biometric schemes aim at generating secure biometric
    templates by combining user specific tokens, such as password, stored
    secret or salt, along with biometric data. This type of transformation
    is constructed as a composition of a biometric transformation with a
    feature extraction algorithm. The security requirements of cancelable
    biometric schemes concern the irreversibility, unlinkability and
    revocability of templates, without losing in accuracy of
    comparison. While several schemes were recently attacked regarding
    these requirements, full reversibility of such a composition in order
    to produce colliding biometric characteristics, and specifically
    presentation attacks, were never demonstrated to the best of our
    knowledge. In this paper, we formalize these attacks for a traditional
    cancelable scheme with the help of integer linear programming (ILP)
    and quadratically constrained quadratic programming (QCQP).  Solving
    these optimization problems allows an adversary to slightly alter its
    fingerprint image in order to impersonate any individual. Moreover, in
    an even more severe scenario, it is possible to simultaneously
    impersonate several individuals. }

\onecolumn \maketitle \normalsize \setcounter{footnote}{0} \vfill

\input{Introduction}

\input{Background}

\input{Attack_desc}

\input{Attack_eval}

\input{Conclusion}

\section*{Acknowledgement}
The authors acknowledges the support of the French Agence Nationale de la Recherche (ANR), under grant ANR-20-CE39-0005 (project PRIVABIO).

\bibliographystyle{apalike}
{\small
    \bibliography{biblio}}

\end{document}

%% file: Introduction.tex
\section{Introduction}

Biometric authentication is more and more used in daily life and are
commonly integrated on many smart objects and devices, \emph{e.g.}, computer, smartphone,
USB drive, passport.  Since biometrics is more convenient and quicker
to use, and biometric characteristics cannot be lost or forgotten,
biometric authentication solutions are in general preferred over their
password or physical token counterparts.
Despite their many advantages, biometric solutions are not exempt from vulnerabilities.
As biometric-based technologies are deployed at a larger scale, centralized biometric databases and devices become natural targets in cyber attacks. These cyber attacks have the potential to be harmful on the long term if they lead to the theft of biometric data. Therfore, a biometric data may actually be vulnerable to impersonation attacks and privacy leakage.

Several criteria essential to biometric authentication systems have
been identified in ISO/IEC 24745~\cite{ISO24745} and ISO/IEC
30136~\cite{ISO30136}:  Irreversibility, unlinkability, revocability
and performance preservation of templates.
\begin{itemize}
      \item \emph{Irreversibility} prevents
            from finding the original person’s biometric data from the protected
            template.
            \vspace*{2 pt}
      \item \emph{Unlinkability} prevents cross-matching attacks or, in
            other words, the linkability between two digital identities,
            \emph{i.e.}, two biometric templates.
            \vspace*{2 pt}
      \item \emph{Revocability} requires the scheme
            to be able to generate new protected templates in case of
            compromission of the biometric database.
            \vspace*{2 pt}
      \item The last criteria,
            \emph{performance preservation}, stipulates that recognition accuracy of
            protected templates should not be degraded compared to the original
            data.
            \vspace*{2 pt}
\end{itemize}
Fulfilling this set of criteria is now necessary to comply with
the \emph{privacy}
principles
of the GDPR.

Faced with the mentioned vulnerabilities and requirements, the
community has proposed primitives dedicated to biometrics, so called
biometric template protection (BTP) schemes.  Examples of such
primitives include cancelable biometrics
(see~\cite{JinLG04,robusthash}), biometric cryptosystems (\emph{e.g.},
fuzzy vault~\cite{FuzzyVault}, fuzzy extractors~\cite{FuzzyExtractor}),
and hybrid
biometrics~\cite{BrChKi08,JaNa12}.  In this paper, we focus on
cancelable biometrics (CB) which is an example of BTP scheme claimed
to meet the four criterias. For more details on BTP schemes, the reader
is referred to two surveys~\cite{Survey-2015}
and~\cite{Survey-2016}. In CB, a biometric template is computed
through a process where the inputs are biometric data (\emph{e.g.}, biometric
image) of a user and a user specific token (\emph{e.g.}, a random key, seed,
salt, or password).  A CB scheme generally consists of a sequence of
processes (an extraction of features followed by a parameterized
transformation) that produces the biometric templates, and a matcher
to generate a matching score between the templates.  With a CB scheme,
templates can be revoked, changed, and renewed by changing user
specific tokens. Even though user tokens in CB may
be considered as secret, the security of a two-factor authentication
system should not be reduced to a single factor. Cryptanalysis of CB
schemes with strong adversarial models commonly assume that the
attacker knows both the biometric template and token of the user. This
assumption is plausible in practice because a user token may have low
entropy (\emph{e.g.}, a weak password), or it may just have been compromised
by an attacker. This stolen-key scenario is also known as the
stolen-token scenario~\cite{TeohKL08}.

Ratha \emph{et al.}~\cite{rcb01} were the first to introduce CB in the case of
face recognition. Since then, several CB schemes have been proposed, including
the popular Biohashing algorithm~\cite{JinLG04} applied on many modalities such
as fingerprints, face, and iris.
CB schemes offer several advantages
such as efficient implementation, high matching accuracy, and revocability.
However, several attacks on a variety of CB schemes have been proposed: attacks
against privacy by approximating feature vectors or linking several templates
of an individual, and authentication attacks by elevating the false
acceptance rate (FAR).
We refer the reader to~\cite{NaNaJa10,ToKaAzEr16} for attacks on biohashing type
schemes, \cite{qfaf08,lh14}
for attacks using the Attack via Record Multiplicity (ARM) technique,
\cite{LCR13,DJJT19} for attacks using genetic algorithms, as well
as attacks using constrained programming on CB schemes built upon
ranking based hashing~\cite{GKLT20}.

Authentication attacks using genetic algorithms have been proposed in~\cite{DJJT19,critique4}.
Their objective is to find the right parameters for generating fingerprint images in order to elevate FAR rates.
In the case of the fingerprint modality, strategies making use of both hill climbing attacks and genetic algorithms
have also been proposed in~\cite{DBLP:journals/corr/abs-1910-07770,critique1}.

\paragraph{Contributions.} In this paper, we propose reversibility attacks against some projection-based CB
schemes, such as the BioHashing~\cite{JinLG04}. The particularity of
our attacks, as opposed to previous works, is that we reverse the
complete sequence of treatments including the \textit{feature
      extraction} algorithm. This allows us to construct impostor
fingerprint images, thus enabling authentication (or presentation)
attacks.
In our authentication attacks, an adversary, who already has
the knowledge of a user’s specific token and has at least one
fingerprint template of the same user, tries to alter his own fingerprint
image such that the adversary can now use its own altered biometrics
and the stolen token to be falsely authenticated as a legitimate
user.
The considered CB schemes are built upon uniform random
projection (URP) and a feature extractor such as Sobel or Gabor filter. To perform our attacks, we use Integer Linear Programming (ILP)
as well as quadratically constrained quadratic programming
(QCQP). Constrained optimization with linear programs has been
previously used in the cryptanalysis of other schemes;
see~\cite{GKLT20,ToKaAzEr16}.


We can state our results as follows:

\textbf{1) Simple authentication attacks.} A complete reversal
methodology of some projection-based CB schemes, including the
BioHash algorithm, is proposed. The main ideas are to solve an
integer linear program and a quadratically constrained quadratic
program to reverse both the projection and the feature extraction.
The solution provided by a solver (\emph{e.g.}, Gurobi) is a
fingerprint image of the attacker whose the amount of changes is
minimized.  Practical resolutions are provided for tiny synthetic
images.

\textbf{2) One fingerprint image for several impersonations.}  The
first attack is extended to produce a fingerprint image that
impersonates the identity of several users. Our formalized
constrained problems and experimentations on tiny synthetic images
show that an adversary can alter its own fingerprint image to be
authenticated as any of several legitimate users. To reach this
objective, two different attacks are proposed:
\begin{itemize}
      \item The first strategy for the attacker is to collect the pairs of (token,
            template) of the target users to enlarge the set of constraints of a QCQP
            program. The solution sought is a single altered fingerprint image of the
            attacker such that, when combined with the distinct stolen tokens, the
            generated templates match exactly the stolen templates of the respective users.
            Impersonating a large number of target users under this approach
            imposes a due acceptance of a larger number of changes in the
            altered fingerprint image of the attacker.
            \vspace*{2 pt}
      \item The second proposed strategy does not require the knowledge of
            the tokens and consists in generating a template which is an average
            (barycentric) template of the target users. Then, the attacker formalizes a
            set of constraints using this template and her token. She solves it to find
            a fingerprint image as close as possible to her own. If the target
            templates lie in a ball of radius two times the decision threshold (in the
            template space), her altered fingerprint image enables an authentication attack
            for multiple users. In other words, her altered image is a
            ``master print'' for these target users.
            \vspace*{2 pt}
\end{itemize}


\paragraph{Outline.} 
Some background information and the adversarial models are presented in
Section~\ref{BackGround}. Section~\ref{AttackMethod} provides
our simple authentication attacks. Section~\ref{mul-auth} introduces an attack not relying on the knowledge of the passwords.
Then, in
Section~\ref{multi-collisions}, it is shown how to impersonate several users with different
passwords. Finally, experimental evaluations and future works are
discussed in Section~\ref{AttackEval} and Section~\ref{Concl} respectively.


%
%
%
%
%
%
%
%


%% file: Background.tex
\section{Background}\label{BackGround}

Cancelable biometric schemes generate secure biometric templates by
combining user specific tokens such as password with his biometric
data such as fingerprint. The goal is to create templates meeting the
four aforementioned criteria, \emph{i.e.}, irreversible, unlinkable,
and revocable templates, with high accuracy of comparison. Biometric
templates in CB schemes are constructed in two steps: ($i$) \emph{Feature extraction}: A feature vector is derived from a biometric image; ($ii$) \emph{Transformation}: A user specific token is used to transform the user's feature vector to a template.


In the following, we let $(\mathcal{M}_I,D_I)$, $(\mathcal{M}_F,D_F)$ and
$(\mathcal{M}_T,D_T)$ be three metric spaces, where
$\mathcal{M}_I$, $\mathcal{M}_F$ and
$\mathcal{M}_T$ represent the fingerprint image space,
the feature space and the template space, respectively;
and $D_I$, $D_F$ and $D_T$ are
the respective distance functions. Note that $D_I$ and $D_F$ are instantiated
with the Euclidean distance, while $D_T$ is instantiated with the Hamming
distance.

\subsection{Feature Extraction with Sobel Filtering}

Let $\mathcal{U}$ be the set of users of the biometric system. We
identify a user with its biometric characteristic, and define a
function $\mathcal{BC}(\cdot)$ that takes a biometric characteristic
$usr \in \mathcal{U}$ as input, and outputs a digital representation
of biometric data $I$; for instance, the scan image of a fingerprint.
Note that for two different computations of $I=\mathcal{BC}(usr)$ and
$I^\prime=\mathcal{BC}(usr)$ (\emph{e.g.}, at different times, or
different devices), we may have $I\ne I^\prime$ due to the inherent
noise in the measurement of biometric data.

\begin{defin}
    A biometric feature extraction scheme is a pair of deterministic polynomial time algorithms $\Pi:=( E,V)$, where:
    \begin{itemize}
        \item $E$ is the feature extractor of the system, that takes biometric data
              $I$ as input, and returns a feature vector $F \in \mathcal{M}_F$.
              \vspace*{2 pt}
        \item $V$ is the verifier of the system, that takes two feature vectors
              $F=E(I)$, ${F^\prime}=E(I')$, and a threshold $\tau$ as input, and
              returns $True$ if $D(F, {F^\prime}) \leq \tau$, and returns $False$ if
              $D(F, {F^\prime}) > \tau$.
              \vspace*{2 pt}
    \end{itemize}
\end{defin}

\noindent
{\bf Sobel Filter.} An example of feature extraction is
\label{Soso}
the Sobel filtering~\cite{Sobel}. Sobel Filter is usually used for edge detection.
The resulting image is obtained by computing two
convolutions given by the following matrices:

$$  G_1 = \left( \begin{matrix}
            1 & 0 & -1 \\
            2 & 0 & -2 \\
            1 & 0 & -1 \\
        \end{matrix} \right) \,\, \text{and} \,\, G_2 = \left(\begin{matrix}
            1  & 2  & 1  \\
            0  & 0  & 0  \\
            -1 & -2 & -1 \\
        \end{matrix} \right).$$

We denote by $ * $ the operator of convolution and by $I$ the matrix of the image in shades of gray. Note that pixels at the
edges of the image are ignored and their values are set to $0$ in the corresponding matrix $I$. The horizontal and vertical gradients, $G_x$ and
$G_y$, are computed as follows $G_x =  G_1 * I$ and $G_y = G_2 * I$ with $*$ the sign of convolution~\ref{Conv} from~\cite{shapiro2001}.
Then, the matrix of the output image $S$ is computed as $\lVert
    G_x + G_y \rVert_2 $ where $\lVert \cdot \rVert_2$ denotes the
Euclidean norm. However, the norm does not apply in the usual way. In fact,
in this case it applies coordinate by coordinate. For example, the first coordinate of $ S $ is
$ S_ {1,1} = \sqrt[2]{G_{x_ {1,1}}^2 + G_{y_ {1,1}}^2 } $.
\begin{defin}[Convolution $*$]
    \label{Conv}
    The general expression of a matrix convolution is
    $$\left[\begin{array}{cccc}
                x_{11} & x_{12} & \cdots & x_{1n} \\
                x_{21} & x_{22} & \cdots & x_{2n} \\
                \vdots & \vdots & \ddots & \vdots \\
                x_{m1} & x_{m2} & \cdots & x_{mn} \\
            \end{array}\right] *
        \left[ \begin{array}{cccc}y_{11} & y_{12} & \cdots & y_{1n} \\
                y_{21}       & y_{22} & \cdots & y_{2n} \\
                \vdots       & \vdots & \ddots & \vdots \\
                y_{m1}       & y_{m2} & \cdots & y_{mn} \\
            \end{array}\right]$$
    $$= \sum_{i=0}^{m-1}\sum_{j=0}^{n-1} x_{(m-i)(n-j)}y_{(1+i)(1+j)}$$
\end{defin}

\begin{figure}[ht]
    \centering
    \includegraphics[width=.4\textwidth]{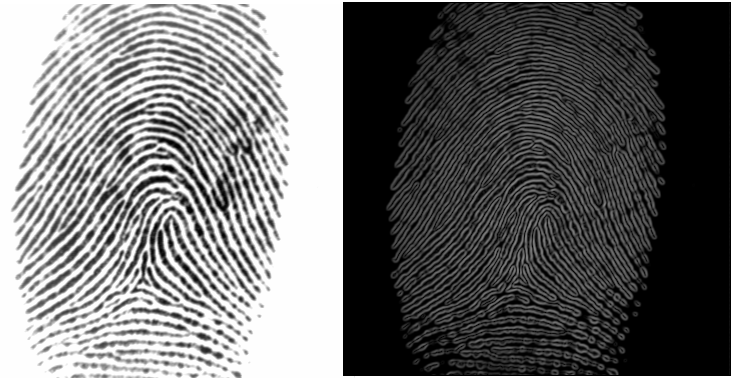}
    \caption{Left: Fingerprint image. Right: Resulting image after
        Sobel filter.}\label{ExemplofSobel}
\end{figure}

Figure~\ref{ExemplofSobel} shows an example of fingerprint input with
its corresponding output by the filter.




%

\subsection{Generation of Templates with URP}
\label{URP}

\begin{defin}
    Let $\mathcal{K}$ be the token (seed) space, representing the set of tokens to be assigned to users. A cancelable biometric scheme is a pair of deterministic polynomial time algorithms $\Xi:=(\mathcal{T}, \mathcal{V})$, where:
    \begin{itemize}
        \item $\mathcal{T}$ is the transformation of the system, that takes a
              feature vector $F \in \mathcal{M}_F$ and the token parameter $P$ as input,
              and returns a biometric template $T=\mathcal{T}(P,F) \in \mathcal{M}_T$.
              \vspace*{2 pt}
        \item $\mathcal{V}$ is the verifier of the system, that takes two biometric
              templates $T$ = $\mathcal{T}(P,F)$,
              ${T^\prime}=\mathcal{T}({P^\prime},{F^\prime})$, and a
              threshold $\tau_T$ as input; and returns
              $True$ if $D_T(T, {T^\prime}) \leq \tau_T$, and
              returns $False$ if $D_T(T, {T^\prime}) > \tau_T$.
              \vspace*{2 pt}
    \end{itemize}
\end{defin}

The attacked CB instantiation, described in Algorithm~\ref{proj}, is
based on a uniform random projection (\textit{URP}). Such a projection
serves as an embedding of a high-dimensional space into a space of
much lower dimension while preserving approximately the distances
between all pairs of points. This type of dimensionality reduction is
characterized by the Johnson–Lindenstrauss
lemma~\ref{JL}~\cite{Johnson1984ExtensionsOL}.
Algorithm~\ref{proj} assumes the second factor,
\emph{i.e.}, the token, is a password and output a \emph{Biometric
    Compressed Vector} (BCV).

\begin{lem}[Johnson-Lindenstrauss]
    \label{JL}
    Given $0<\epsilon < 1$, a set $X$ of $m$ point in $\mathbb{R}^N$,
    and a number $n > 8\left(\dfrac{ln\left(m\right)}{\epsilon^2}\right)$, there is a
    linear map $f:\mathbb{R}^N \mapsto \mathbb{R}^n$ such that for all
    u,v $\in \mathbb{X}$ :
    $$\left(1-\epsilon \right) \left | \left | u-v \right | \right | \leq \left |
        \left | f\left (u\right)-f\left (v\right) \right | \right | \leq
        \left(1+\epsilon \right) \left | \left | u-v \right | \right | $$
\end{lem}
%
%
\begin{algorithm}
    \caption{\sc[URP-Sobel]}\label{proj}
    \hspace*{\algorithmicindent} \textbf{Inputs :} biometric data $I$; token parameter $P$ \\
    \hspace*{\algorithmicindent} \textbf{Output :} $T=(t_1, \ldots, t_m)$ \\
    \hspace*{\algorithmicindent} $T$ is a  BCV (Biometric Compressed Vector)
    \begin{algorithmic}[1]
        \State Apply Sobel filter on $I$ to produce an
        $n$-sized feature vector: $F=(f_1,\dots,f_n)$.
        \State Generate with the token $P$ a family $V$  of $m$
        pseudorandom vectors $V_1,\dots,V_m$ of size $n$ according to a
        uniform law $\mathcal{U}(\left[-0.5,0.5\right])$.
        \State Arrange the family $V$ as a matrix $M$ of size
        $n\times m$.
        \State Compute $T$ as the matrix-vector product $F\times M$.
        \For{$t_i$ in $T$}
        \IfThenElse {$t_i< 0$}{$t_i=0$}{$t_i = 1$}
        \EndFor\\
        \Return $T$
    \end{algorithmic}
\end{algorithm}

%

\begin{rem}
    Biohashing instantiation~\cite{JinLG04} is based on the same type of projection,
    except that an additional step of orthonormalization of the family $V$ by Gram-Schmidt is performed.
    This skipped step affects neither the recognition accuracy nor the feasibility of the attacks.
    However, their running times are reduced. Indeed, experiments over \textit{FVC-2002-DB1} using the URP-Sobel scheme yield a decision threshold at $225$ for
    an EER equal to $0.29\%$. However, in the case of Biohashing, the same experiments yield a decision threshold at $224$ for
    an EER equal to $0.27\%$. Therefore, the recognition accuracy results are pretty similar whether or not orthonormalization is performed.
\end{rem}


\subsection{Attack Models and Objectives}

We perform an authentication attack and, we are able to get access to
this system in the name of the targeted person.

To perform this attack some information are needed:
\begin{itemize}
    \item The password of our target.
          \vspace*{2 pt}
    \item The original biohash of the target.
          \vspace*{2 pt}
    \item Knowledge over the attacked system:
          \vspace*{2 pt}
          \begin{itemize}
              \item How to get the matrix from the password.
                    \vspace*{1.5 pt}
              \item The value of the quantization that was used to create the BCV.
                    \vspace*{2 pt}
          \end{itemize}
\end{itemize}

We show that anybody can perform a simple authentication attack or a one fingerprint image
for several impersonations attack by building a template preimage if he knows the above information.


The informal definitions of \cite{GKLT20} are tailored for the rest of the
paper. Let $I\in \mathcal{M}_I$ be a fingerprint image,
and let $T=\Xi.\mathcal{T}(P,E(I))\in\mathcal{M}_T$ be the template generated
from $I$ and the secret parameter $P$. In our authentication attack, an
adversary is given $T$, $P$, and a threshold value $\tau_T$, and the adversary
tries to find a fingerprint image $I^*\in \mathcal{M}_I$ such that
for $T^*=\Xi.\mathcal{T}(P,E(x^*))$, $T^*$ is exactly the same as $T$,
or $T^*$ is close to $T$ with respect
to the distance function over $\mathcal{M}_T$ and the threshold value $\tau_T$.
In this case, we say that $I^*$ is a $\tau_T$-nearby-template preimage (or
simply a nearby-template preimage, when $\tau_B$ is clear from the context) of
the template $T$.

A strategy for the adversary which have stolen the secret parameter $P$
is to alter her fingerprint image $I_A$ such that $P$ along with her extracted
feature vector $F_A$ enable the generation of the exact template
$T$. This motivates the notion of \textit{template fingerprint preimage} defined
below.

\begin{defin}[Template fingerprint preimage]
    Let $I\in \mathcal{M}_I$ be a fingerprint image,
    and $T=\Xi.\mathcal{T}(P,\Pi.E(I))\in\mathcal{M}_T$ a template for some secret parameter
    $P$. A {\it template preimage} of $T$ with respect to $P$
    is a fingerprint image $I^*$ such that
    $T=\Xi.\mathcal{T}(P,\Pi.E(I^*))$.
\end{defin}

Another authentication attack consists in generating a fingerprint image that
yields the exact templates of two distinct users with their
corresponding stolen tokens. More formally, we have the following definition:

\begin{defin}[Two-template fingerprint preimage]
    Let $I_1$ and $I_2\in \mathcal{M}_I$ be two fingerprint images of distinct users,
    and two templates $T_1=\Xi.\mathcal{T}(P_1,\Pi.E(I_1))\in\mathcal{M}_T$ and
    $T_2=\Xi.\mathcal{T}(P_2,\Pi.E(I_2))\in\mathcal{M}_T$ for distinct secret
    parameters $P_1$ and $P_2$. A {\it two-template preimage} of the pair
    $(T_1, T_2)$ with respect to the pair $(P_1, P_2)$ is a fingerprint image
    $I^*$ such that $T_1=\Xi.\mathcal{T}(P_1,\Pi.E(I^*))$ and
    $T_2=\Xi.\mathcal{T}(P_2,\Pi.E(I^*))$.
\end{defin}

To capture the case of multi-collisions, this last definition can be
generalized to a notion of a $n$-template fingerprint preimage.

\begin{defin}[$n$-template fingerprint preimage]
    Let \\ $I_1,\dots,I_n\in \mathcal{M}_I$ be $n$ fingerprint images of distinct users,
    and $n$ templates $T_i=\Xi.\mathcal{T}(P_i,\Pi.E(I_i))\in\mathcal{M}_T$
    for distinct secret parameters $P_i$ $\forall i \in \lbrace 0,\dots,n \rbrace$.
    A {\it n-template preimage} of $(I_1,\dots,I_n)$ with respect to $(P_1,\dots ,P_n)$ is a fingerprint image
    $I^*$ such that: $$\forall i \in \lbrace 0,\dots,n \rbrace, T_i=\Xi.\mathcal{T}(P_i,\Pi.E(I^*)).$$
\end{defin}

%% file: Attack_desc.tex
\section{Simple Authentication Attack}
\label{AttackMethod}

\subsection{Overview}

There are two ways of performing this attack. The first one includes two steps described in Section~\ref{Method1}.
First, given an attacker feature vector, we seek the slightest modification of it such that its transformation by $\Xi$ yields exactly the template of the victim.
Then, using the filter constraints of the convolution, we seek the slightest variation of the attacker's image such that the filtering of this variation produces exactly the modified feature vector.
The second approach described in Section~\ref{Method2} consists in generating all constraints at once and directly generating the modified
attacker's image.

\subsubsection{First Approach} \label{Method1}

The attack takes as input the following parameters:
\begin{itemize}
  \item The target's password ($P_t$).
        \vspace*{2 pt}
  \item The target's template ($T_t$).
        \vspace*{2 pt}
  \item The attacker's image ($I_A$).
        \vspace*{2 pt}
\end{itemize}

This attack computes and uses following information:
\begin{enumerate}
  \item Attacker's feature ($F_A$).
        \vspace*{2 pt}
  \item Modified attacker's feature ($F_A'$).
\end{enumerate}

The output is a modified attacker's image $X$ which matches the target
template.

First, the attacker uses $I_A$ to compute $F_A$ using filter. Then, with
$P_t$ and $T_t$, the attacker modifies image's feature to match the
target template $F_A'$. As described in Section~\ref{GCOF}, it is done by
solving an under-constraint linear system and seeking the nearest
modified feature which matches the target template. After that, using $F_A'$ and
$I_A$, the attacker modifies its image to match the modified feature. As
described in Section~\ref{GPAFE}, it is done by solving an
under-constraint quadratic system and seeking the nearest modified image which
matches the feature.

Figure~\ref{Process} gives an overview of this first method step by
step, where inputs are in circles and different steps in boxes.

\tikzstyle{decision} = [diamond, draw, fill=blue!20,
text width=4.5em, text badly centered, node distance=2.5cm, inner sep=0pt]
\tikzstyle{block} = [rectangle, draw, fill=blue!20,
text width=7em, text centered, rounded corners, minimum
height=4em, minimum width=7em]
\tikzstyle{block2} = [rectangle, draw, fill=blue!20,
text width=11em, text centered, rounded corners, minimum
height=4em, minimum width=10em]
\tikzstyle{line} = [draw, very thick, color=black, -latex']
\tikzstyle{cloud} = [draw, ellipse,fill=red!20, node distance=2.5cm,
minimum height=2em]
\tikzstyle{decision answer}=[near start,color=black]

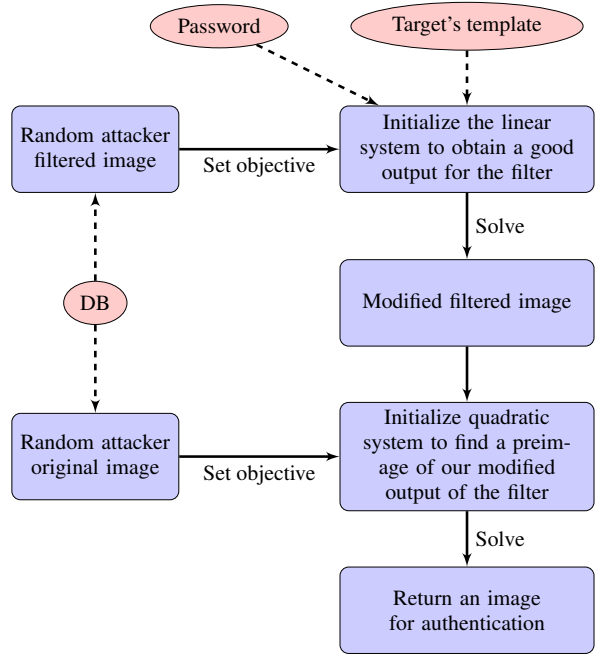
\begin{figure}[tb]
  \resizebox{.49\textwidth}{!}{%
    \begin{tikzpicture}[node distance = 2cm, auto]
      \node [cloud] (BCV) {Target's template};
      \node [cloud, left of=BCV, node distance=4cm] (pass) {Password};

      \node [block2, below of=BCV] (sys) {Initialize the linear system to obtain a
        good output for the filter};
      \node [block,left of=sys, node distance=6cm] (Filtered) {Random
        attacker filtered image};
      \node [block2,below of=sys, node distance=2.5cm] (Mod) {Modified
        filtered image};
      \node [block2,below of=Mod, node distance=2.5cm] (Init) {Initialize
        quadratic system to find a preimage of our modified output of the filter};
      \node [block2,below of=Init, node distance=2.5cm] (Return) {Return
        an image for authentication};

      \node [cloud, left of=Mod, node distance=6cm] (data) {DB};
      \node [block, below of=data, node distance=2.5cm] (ori) {Random
        attacker original image};

      \path [line,dashed] (pass) -- (sys);
      \path [line,dashed] (BCV) -- (sys);
      \path [line] (Filtered) --  node [below] {Set objective} (sys);
      \path [line] (sys) --  node [] {Solve} (Mod);
      \path [line] (Mod) -- (Init);
      \path [line] (Init) --  node [] {Solve} (Return);
      \path [line,dashed] (data) --  (Filtered);
      \path [line,dashed] (data) --  (ori);
      \path [line] (ori) --  node [below] {Set objective} (Init);
    \end{tikzpicture}
  }
  \caption{Principle of the attack's first approach.}\label{Process}
\end{figure}

\subsubsection{Second Approach} \label{Method2}
The attack takes as input the same parameters ($P_t$, $T_t$ and $I_A$).
The output is a modified attacker's image $X$ which matches the target
template.

The main idea is to merge both steps described in
Section~\ref{Method1}. A unique constrained quadratic system is solved 
to find the nearest modified image which matches the template (see Figure~\ref{Process2}).  

\tikzstyle{decision} = [diamond, draw, fill=blue!20,
text width=4.5em, text badly centered, node distance=2.5cm, inner sep=0pt]
\tikzstyle{block} = [rectangle, draw, fill=blue!20,
text width=7em, text centered, rounded corners, minimum
height=4em, minimum width=7em]
\tikzstyle{block2} = [rectangle, draw, fill=blue!20,
text width=11em, text centered, rounded corners, minimum
height=4em, minimum width=10em]
\tikzstyle{line} = [draw, very thick, color=black, -latex']
\tikzstyle{cloud} = [draw, ellipse,fill=red!20, node distance=2.5cm,
minimum height=2em]
\tikzstyle{decision answer}=[near start,color=black]

\begin{figure}[tb]
  \resizebox{.49\textwidth}{!}{%
    \begin{tikzpicture}[node distance = 2cm, auto]
      \node [cloud] (BCV) {Target's template};
      \node [cloud, left of=BCV, node distance=4cm] (pass) {Password};

      \node [block2, below of=BCV] (sys) {Initialize the quadratic system};
      \node [block,left of=sys, node distance=6cm] (ori) {Random
        attacker image};
      \node [block2,below of=sys, node distance=2.5cm] (Return) {Return
        an image for authentication};

      \node [cloud, left of=Mod, node distance=6cm] (data) {DB};

      \path [line,dashed] (pass) -- (sys);
      \path [line,dashed] (BCV) -- (sys);
      \path [line] (Filtered) --  node [below] {Set objective} (sys);
      \path [line] (sys) --  node [] {Solve} (Return);
      \path [line,dashed] (data) --  (ori);
    \end{tikzpicture}
  }
  \caption{Principle of the attack's second approach.}\label{Process2}
\end{figure}
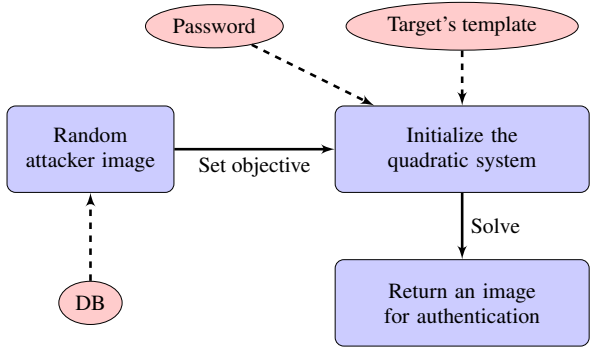

\subsection{Program Formulation for the Two-Phase Approach}

As explained, we proceed in two steps.

\subsubsection{Getting a Correct Output for the Filter}
\label{GCOF}

For this part, we assume that we are after the filter. We see how to
inverse the filter later.

We want to reverse target's template by using the
password. To do that, let $X=(x_0,\dots,x_n)$, $M$ the projection
matrix derivated from target's password and $f$ the quantization function which
takes $X M$ to create a binary template.

We know the projection matrix and the image we need to get for the
target client. Thus, one can seek to calculate a pre-image of the
projected vector by solving a system under constraints.

\begin{rem}
  This attack works for many projections system such as Biohash.
\end{rem}

Let us write it more formally. Let $T=(t_1,\dots,t_m)$ the biometric
template, $n$ the size of BCV and
\[ M = \begin{bmatrix}
    a_{1,1} & \dots  & a_{0,n} \\
    \vdots  & \ddots & \vdots  \\
    a_{m,0} & \dots  &
    a_{m,n}                    \\\end{bmatrix}.\]

Let $\mathcal{K}_1$ be all indices where the template is equal to $0$ and $\mathcal{K}_2$ all
other indices. So, we seek a solution to the following system:

\begin{equation}\label{EQ2}
  \begin{cases}
    X M_i < 0                ,
    \forall i \in \mathcal{K}_1                                                        \\
    X M_i \ge 0              , \forall j \in \mathcal{K}_2                             \\
    x_i \in \mathbb{R}^+ , \forall i \in \left(\mathcal{K}_1 \cup \mathcal{K}_2\right) \\
  \end{cases}
\end{equation}

With $M_i$ the $i$-th column of $M$.
We seek to minimize the distance between $F$ and
$F_A$. By doing so, the attacker can be authenticated by
modifying the smallest number of information of his own biometric
feature vector.

This part of the attack solves the following problem. By taking $F_A =
  (o_1, \dots, o_n)$ the attacker's biometric feature, $M$ the projection
matrix we have:

\begin{itemize}
  \item Minimize: $\|X-F_A\|^2$
        \vspace*{2 pt}
  \item Under the following constraints:
        \begin{equation}
          \begin{cases}
            X M_i < 0                ,
            \forall i \in \mathcal{K}_1                                                        \\
            X M_i \ge 0              , \forall j \in \mathcal{K}_2                             \\
            x_i \in \mathbb{R}^+ , \forall i \in \left(\mathcal{K}_1 \cup \mathcal{K}_2\right) \\
          \end{cases}
        \end{equation}
        \vspace*{2 pt}
\end{itemize}
With $M_i$ the $i$-th column of $M$.
\subsubsection{Get a Preimage to Avoid Filter Effect}
\label{GPAFE}

The filter leads to a loss of information. But we can write a
quadratic system to create a collision and get a correct preimage.
Let the image matrix be \[ I = \begin{bmatrix}
    o_{0,0}               & \dots  & o_{0,\text{width-1}} \\
    \vdots                & \ddots & \vdots               \\
    o_{\text{length-1},0} & \dots  &
    o_{\text{length-1},\text{width-1}}                    \\\end{bmatrix}\]

Applying the filter to that formal matrix yields a new matrix $D$ which has
quadratic components. But, we know that $D$ must be equal to
$F_A$. Thus, we can solve a quadratic system with
$\left(\text{length}\times\text{width}\right)$ equations and
$\left(\text{length}\times\text{width}\right)$ variable to find a
preimage.

Let $I_A = (o_{i,j})$ denote the attacker's original image, $F_A =
  (a_{i,j})$ its modified feature, $I' = (x_{i,j}')$ the modified
original image and $X = (x_{i,j})$ its augmented form. We consider
the augmented form as the original matrix where zeroes are added all
around the matrix to compute the convolution without overflowing.

In the case of Sobel filter, we solve the following problem:
\begin{itemize}
  \item Minimize: $\sum\limits_{i,j}^{} \left( o_{i,j} - x_{i,j} \right )^2$\\
        \vspace*{2 pt}
  \item Subject to the following constraints:
        \begin{equation}
          \begin{cases}
            \alpha_{i,j} =  x_{(i-1,j-1)}+2x_{(i,j-1)}+x_{(i+1,j-1)}                          \\
            \phantom{\alpha_{i,j} =}                -x_{(i-1,j+1)}-2x_{(i,j+1)}-x_{(i+1,j+1)} \\
            \beta_{i,j} =   x_{(i-1,j-1)}+2x_{(i-1,j)}+x_{(i-1,j+1)}                          \\
            \phantom{\beta_{i,j} =}               -x_{(i+1,j-1)}-2x_{(i+1,j)}-x_{(i+1,j+1)}   \\
            a_{i,j}^2 =    \alpha_{i,j}^2+\beta_{i,j}^2 , \forall (i,j)                       \\
            x_{i,j} = 0 \text{ if }  i=0 \text{ or } i=length + 1                             \\
            x_{i,j} = 0 \text{ if }  j=0 \text{ or } j=width  + 1                             \\
            x_{i,j} \in    \left\llbracket 0,255\right\rrbracket ,\forall (i,j)               \\
          \end{cases}
        \end{equation}
        \vspace*{2pt}
\end{itemize}

Using the notations of Section~\ref{Soso}, we obtain:

\begin{itemize}
  \item Minimize: $\|X-I_A\|^2$
        \vspace*{2pt}
  \item Under the following constraints:
        \begin{equation}
          \begin{cases}
            (F_A)^2 = \left[(G_1 * X)^2 + (G_2 * X)^2 \right] \\
            x_{i,j}   \in \left\llbracket 0,255\right\rrbracket,  \forall (i,j)
          \end{cases}
        \end{equation}
        \vspace*{2 pt}
\end{itemize}

\subsection{Formulation as a Single Program}
\label{merge}

Yet another method is to merge both systems to create a
new quadratic system. In this case, we avoid some problems such as
having an intermediate feature vector which is not in the range of the
filter function.

Assume that $I_A = (o_{i,j})_{n\times m}$ is the attacker's original image, $I' =
  (x_{i,j}')_{n\times m}$ the modified original image and $X = (x_{i,j})_{n\times m}$ its
augmented form. Let $\mathcal{K}_1$ be all indices where the template
is equal to $0$ and $\mathcal{K}_2$ all other indices. Let $M =
  (a_{i,j})_{(n*m) \times \ell}$ be the
projection matrix. Let $Y_{flat}$ be the flattened form of the matrix $Y$
where rows are concatenated in a single vector.

Thus, using the notations from the sections~\ref{Method1} and \ref{Method2}
we define the following problem for Sobel filter:

\begin{itemize}
  \item Minimize: $\|X-I_A\|^2$
        \vspace*{2 pt}
  \item Under the following constraints:
        \begin{equation}
          \begin{cases}
            Y^2 =                        \left[(G_1 * X)^2 + (G_2 * X)^2 \right] \\
            Y_{flat} M_i < 0                    , \forall i \in \mathcal{K}_1    \\
            Y_{flat} M_j \ge 0                 ,  \forall j \in \mathcal{K}_2    \\
            x_{i,j} \in \left\llbracket 0,255\right\rrbracket, \forall (i,j)
          \end{cases}
        \end{equation}
        \vspace*{2 pt}
\end{itemize}

Where the squaring stands for the coordinate by coordinate squaring (\textit{i.e.} Hadamard squaring)
and $M_i$ the $i$-th column of $M$.



\section{Multiple Authentications Attack}
\label{mul-auth}

The goal of this attack is to find an image that can impersonate
several victims. The attacker computes
an image whose derived template is kind of a barycenter of all the targeted templates
This is possible only if these templates are not too
far with respect to a threshold. A specificity of this attack is
that the passwords of the victims are not required.

\subsection{Overview}
\label{MMM}
Let $\mathcal{\mu}$ denote the set of the victims. The attack takes as input the following parameters:
\begin{enumerate}
  \item The target's templates $(T_t)_{i \in \mathcal{\mu}}$.
        \vspace*{2 pt}
  \item The attacker's image ($I_A$).
        \vspace*{2 pt}
  \item The attacker's password ($P_A$).
        \vspace*{2 pt}
  \item The value of $\epsilon$ the decision threshold.
        \vspace*{2 pt}
\end{enumerate}

The output is a modified attacker’s image $X$ which
matches the modified template.

First, with respect to all the targeted templates, we seek a template $T$
such that they are in a ball centered in $T$ and of radius $\epsilon$.
If a center\footnote{Several centers may exist and, with more than $2$ templates the
  existence of at least one center is not ensured.} does not exist, a subset of the targeted
templates for which the center exists is considered.

Then, a quadratic system and a function to minimize can be built as
explained in Section~\ref{Multi-at}.  Thus, solving this problem gives
us the modified image for multiple authentications with the same
password.  We present an overview of this attack in
Figure~\ref{Process3}.

\tikzstyle{decision} = [diamond, draw, fill=blue!20,
text width=4.5em, text badly centered, node distance=2.5cm, inner sep=0pt]
\tikzstyle{block} = [rectangle, draw, fill=blue!20,
text width=7em, text centered, rounded corners, minimum
height=4em, minimum width=7em]
\tikzstyle{block2} = [rectangle, draw, fill=blue!20,
text width=11em, text centered, rounded corners, minimum
height=4em, minimum width=10em]
\tikzstyle{line} = [draw, very thick, color=black, -latex']
\tikzstyle{cloud} = [draw, ellipse,fill=red!20, node distance=2.5cm,
minimum height=2em]
\tikzstyle{decision answer}=[near start,color=black]

\begin{figure}[tb]
  \resizebox{.49\textwidth}{!}{%
    \begin{tikzpicture}[node distance = 2cm, auto]
      \node [cloud] (BCV) {Target's templates};
      \node [cloud, left of=BCV] (seuil) {$\epsilon$};
      \node [block, below of=BCV] (NT) {Compute the center template};
      \node [cloud, left of=NT, node distance=4cm] (pass) {Attacker's Password};

      \node [block2, below of=NT] (sys) {Initialize the quadratic system};
      \node [block,left of=sys, node distance=6cm] (ori) {Random
        attacker image};
      \node [block2,below of=sys, node distance=2.5cm] (Return) {Return
        an image for authentication};

      \node [cloud, left of=Return, node distance=6cm] (data) {DB};

      \path [line,dashed] (BCV) -- (NT);
      \path [line,dashed] (seuil) -- (NT);
      \path [line,dashed] (NT) -- (sys);
      \path [line,dashed] (pass) -- (sys);
      \path [line] (ori) --  node [below] {Set objective} (sys);
      \path [line] (sys) --  node [] {Solve} (Return);
      \path [line,dashed] (data) --  (ori);
    \end{tikzpicture}
  }
  \caption{Principle of the multiple authentications attack.}\label{Process3}
\end{figure}
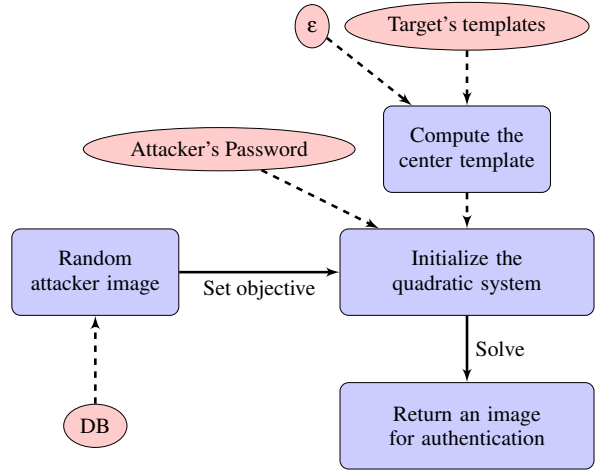

\subsection{Program Formulation}
\label{Multi-at}

Let $M$ be the projection matrix and $T$ the template at the center of
the ball as defined in Section~\ref{MMM}. Assume that $\mathcal{K}_1$ is the
list of all indices where $T_i$ is equal to $0$ and $\mathcal{K}_2$ is the list of the remaining
indices. The other notations are the same as in Section~\ref{merge}. The problem
can be defined like this:
\begin{itemize}
  \item Minimize: $\|X-I_A\|^2$
        \vspace*{2 pt}
  \item Under the following constraints where $M_k$ is the $k$-th
        column of $M$:
        \begin{equation}
          \begin{cases}
            Y^2 =    \left[(G_1 * X)^2 + (G_2 * X)^2 \right]                     \\
            Y_{flat}  M_j < 0                      , \forall j \in \mathcal{K}_1 \\
            Y_{flat}  M_k  \ge 0                  , \forall k \in \mathcal{K}_2  \\
            x_{i,j}  \in \left\llbracket 0,255\right\rrbracket , \forall (i,j)
          \end{cases}
        \end{equation}
        \vspace*{2 pt}
\end{itemize}

With $M_i$ the $i$-th column of $M$.

\section{Multiple Collisions Attack}\label{multi-collisions}

In this attack, the attacker knows the templates and passwords of the
victims. Then, his goal is to use all these information to generate one
image that allows her to impersonate
all the victims using their own password.

\subsection{Overview}

\label{Mam}
The attack takes as input the following parameters:
\begin{enumerate}
  \item The target's templates $(T_t)_{i \in \mathcal{\mu}}$.
        \vspace*{2 pt}
  \item The attacker's image ($I_A$).
        \vspace*{2 pt}
  \item The target's passwords $(P)_{i \in \mathcal{\mu}}$.
        \vspace*{2 pt}
\end{enumerate}

The output is a modified attacker’s image $X$ which
matches all templates for the corresponding password.

\begin{figure}[tb]
  \resizebox{.49\textwidth}{!}{%
    \begin{tikzpicture}[node distance = 2cm, auto]
      \node [cloud] (BCV) {Targets's templates};
      \node [cloud, left of=BCV, node distance=4cm] (pass) {Passwords};

      \node [block2, below of=BCV] (sys) {Initialize the quadratic system};
      \node [block,left of=sys, node distance=6cm] (ori) {Random
        attacker image};
      \node [block2,below of=sys, node distance=2.5cm] (Return) {Return
        an image for authentication};

      \node [cloud, left of=Mod, node distance=6cm] (data) {DB};

      \path [line,dashed] (pass) -- (sys);
      \path [line,dashed] (BCV) -- (sys);
      \path [line] (Filtered) --  node [below] {Set objective} (sys);
      \path [line] (sys) --  node [] {Solve} (Return);
      \path [line,dashed] (data) --  (ori);
    \end{tikzpicture}
  }
  \caption{Principle of the attack's second approach.}\label{Process4}
\end{figure}
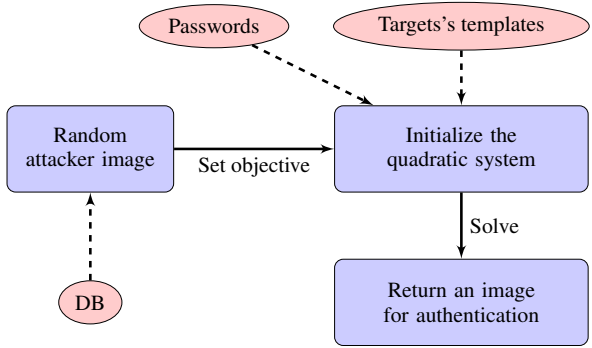

We define a quadratic system and a function to minimize as explained in
Section~\ref{Multi}.  Thus, solving this problem gives us the modified
image for multiple authentications for each password.
An overview of this attack is depicted in Figure~\ref{Process4}.

\subsection{Program Formulation}
\label{Multi}

Let $M_i$ be the projection matrix for the $i$-th user. Assume that $(\mathcal{K}_1)_i$
is the list
of all indices where $(T_t)_i$ is equal to $0$ and $(\mathcal{K}_2)_i$ all other
indices.  The other notations are the same as in Section~\ref{merge}.  The problem
can be defined like this:
\begin{itemize}
  \item Minimize: $\|X-I_A\|^2$
        \vspace*{2 pt}
  \item Under the following constraints where $(M_i)_j$ is the $j$-th column
        of $M_i$:
        \begin{equation}
          \begin{cases}
            Y^2 =                        \left[(G_1 * X)^2 + (G_2 * X)^2 \right]                                \\
            Y_{flat}  (M_i)_j < 0                 ,\forall i \in \mathcal{\mu}, \forall j \in (\mathcal{K}_1)_i \\
            Y_{flat}  (M_i)_k \ge 0              , \forall i \in \mathcal{\mu}, \forall k \in (\mathcal{K}_2)_i \\
            x_{i,j} \in \left\llbracket 0,255\right\rrbracket , \forall (i,j)
          \end{cases}
        \end{equation}
        \vspace*{2 pt}
\end{itemize}

As matrices $M_i$ are fully random, the
probability of them forming an indexed family of linearly dependent vectors is negligible, thus making the system solvable.
Assume that
$L(V_1,\dots,V_k)$ is the event that $(V_1,\dots,V_k)$ is an indexed family of linearly independent vectors, with $n$
the size of vector and $\eta$ the number of precision bits for our numbers.
It can be shown that
$$P(L(V_1,\dots,V_k)) = \dfrac{\prod_{i=2}^{k} 2^{\eta(n-i+1)}-1}{\prod_{i=2}^{k} 2^{\eta(n-i+1)}}.$$

Since this probability is near $1$, the usurpation of $\left \lfloor \dfrac{n}{w} \right \rfloor$ persons
with $w$ the size of the template is a likely event.

\begin{rem}
  A variant of this attack could be achieved without the users' passwords.
  The attacker just has to replace the passwords of the victims by distinct random strings.
  Thus, she obtains an image that allows her to impersonate several people. She merely
  chooses one victim by using its corresponding string. However, it may be possible that
  its string lead to an infeasible model and so another must be chosen.
\end{rem}

%% file: Attack_eval.tex
\section{Reversibility Attack Evaluation}\label{AttackEval}


We evaluate the impact of our authentication
attack with the second approach (\ref{Method2}) through our Python implementation.
The Gurobi Python interface (Gurobi $9.1.2$)
is used to solve the non-convex quadratically constrained programs, on a computer running on
Debian 11, with an EPYC 7F72 dual processor (48 cores) and 256GB of RAM.
The focus is only done on the results of this attack because its practicality
implies the practicality of the others.

We have launched resolutions of the constrained programs $50$ times, each with a time limit of $150$ seconds.
Table~\ref{run_time} reports the running times for the different settings along with the amount of changes
done in the attacker fingerprint, by means of the Euclidian distance.

In Table~\ref{run_time}, we remark that with a 4$\times$4-pixel image and a $50$-bit template,
the hard cap of $150$ seconds starts to be insufficient to solve the system and optimize the criterion.
However, the experiments are encouraging given that we face an NP-hard problem~\cite{QP-NP-HARD}.
By setting the hard cap to $500$ seconds, we are able to solve the system with a 10$\times$10-pixel
image and a better ratio amount of changes over image size.

\begin{table}[h]
    \begin{center}
        \resizebox{.49\textwidth}{!}{%
            \begin{tabular}{|c|c|c|c|}
                \hline
                Image Size   & Template Size & Mean Distance & Mean Time (s) \\
                \hline\hline
                $2 \times 2$ &               & $58$          & $0.04$        \\
                $2 \times 3$ &               & $65$          & $27.92$       \\
                $3 \times 3$ & $20$          & $71$          & $120.18$      \\
                $4 \times 3$ &               & $95$          & $135.56$      \\
                $4 \times 4$ &               & $152$         & $140.83$      \\
                \hline
                $2 \times 2$ &               & $81$          & $1.58$        \\
                $2 \times 3$ &               & $83$          & $77.24$       \\
                $3 \times 3$ & $30$          & $76$          & $129.36$      \\
                $4 \times 3$ &               & $102$         & $138.0$       \\
                $4 \times 4$ &               & $153$         & $143.18$      \\
                \hline
                $2 \times 2$ &               & $119$         & $0.12$        \\
                $2 \times 3$ &               & $102$         & $33.1$        \\
                $3 \times 3$ & $40$          & $121$         & $144.0$       \\
                $4 \times 3$ &               & $133$         & $146.81$      \\
                $4 \times 4$ &               & $168$         & $146.60$      \\
                \hline
                $2 \times 2$ &               & $99$          & $0.14$        \\
                $2 \times 3$ &               & $117$         & $32.76$       \\
                $3 \times 3$ & $50$          & $133$         & $150.0$       \\
                $4 \times 3$ &               & $144$         & $146.67$      \\
                $4 \times 4$ &               & $177$         & $150.0$       \\
                \hline
            \end{tabular}}
    \end{center}
    \caption{Summary of the experiments.}
    \label{run_time}
\end{table}



%% file: Conclusion.tex
\section{Concluding Remarks}\label{Concl}

In this paper, we present several authentication attacks on a popular
CB scheme consisting in a composition of a kernel-based filter
with a projection-based transformation, in the stolen token scenario. Their
particularity is to completely reverse a CB scheme.
to impersonate any or several users.
To the best of our knowledge, this is the first time that attacks
are conducted on a complete chain of treatments, including a non-linear filter. The proposed methodology is
to formalize the attacks as
constrained optimization problems. As long as the attacker
has access to one or several templates with the corresponding passwords,
our attacks can be performed. In addition, we present two ways for the attacker
to impersonate several legitimate persons
Some attacks proposed do not need any token from the clients.
Our practical experiments show that the modification of the attacker's
image is minimal over small images. The next step is to perform these
attacks on larger images and look for the limit of the number of people
that can be impersonated at the same time.
Future work will be focused on
finding optimizations and relaxations of the systems to ensure the scaling
of our attacks.
